\documentclass{article}
\usepackage{verbatim}
\usepackage{arxiv}
\usepackage{setspace}
\usepackage{array,multirow}
\usepackage{natbib}
\usepackage{graphicx}
\usepackage{amsmath}
\usepackage[
  locale = DE 
]{siunitx}
\usepackage[utf8]{inputenc} 
\usepackage[T1]{fontenc} 
\usepackage{hyperref} 
\usepackage{url}
\usepackage{booktabs}
\usepackage{amsfonts}
\usepackage{nicefrac} 
\usepackage{microtype} 
\newtheorem{theorem}{Theorem}

\newcommand\independent{\protect\mathpalette{\protect\independenT}{\perp}}
\def\independenT#1#2{\mathrel{\rlap{$#1#2$}\mkern2mu{#1#2}}}
\title{Sensitivity analysis for incomplete data \emph{via} unmeasured confounding}

\author{
  Heng Chen\\
  Department of Biostatistics\\ 
  Gilead Sciences, Inc.\\
  \texttt{Henry.Chen10@gilead.com} \\
   \And
 Daniel F. Heitjan%\thanks{Use footnote for providing further information about author (webpage, alternative address)---\emph{not} for acknowledging funding agencies.}
 \\
 Department of Statistical Science, Southern Methodist University\\
 Department of Population \& Data Sciences, University of Texas Southwestern Medical Center\\
  \texttt{dheitjan@smu.edu}\\
}
\begin{document}
\maketitle
%\doublespacing
\begin{abstract}
We present a method to analyze sensitivity of frequentist inferences to potential nonignorability of the missingness mechanism.  Rather than starting from the selection model, as is typical in such analyses, we assume that the missingness arises through unmeasured confounding.  Our model permits the development of measures of sensitivity that are analogous to those for unmeasured confounding in observational studies.  We define an index of sensitivity, denoted \emph{MinNI}, to be the minimum degree of nonignorability needed to change the mean value of the estimate of interest by a designated amount.  We apply our model to sensitivity analysis for a proportion, but the idea readily generalizes to more complex situations.
\end{abstract}


\section{Introduction}
Rubin has elucidated the role of the missingness mechanism in extracting frequentist inferences from incomplete data. \citep{rubin1976inference} The idea is to compare the distribution of the variables that are observed, conditional on the observed missingness indicators, to the marginal distribution of these same variables ignoring the missingness mechanism.  The condition \emph{missing completely at random (MCAR)} is sufficient to guarantee that these distributions are identical, and therefore that the missingness mechanism is \emph{ignorable}. \citep{little2014statistical} Briefly, MCAR requires that the conditional probability of the observed missingness indicators given the notional complete data is independent of the value of the complete data.  Heitjan has extended this analysis to the coarse data model, where MCAR generalizes to \emph{coarsened completely at random}. \citep{heitjan1994ignorability,heitjan1997ignorability}

The Rubin approach begins with a \emph{selection model}; that is, it parameterizes the joint distribution of the variable of interest $Y$ and the missingness indicator $G$ as the product of the marginal distribution of $Y$ times the conditional distribution of $G$ given $Y$, denoting the latter the \emph{missingness mechanism}.  One can then identify ignorability conditions as restrictions on the parameters of the missingness mechanism.  It is not possible to estimate the parameters of this joint model without strong assumptions. \citep{diggle1994informative} An alternative, less ambitious, approach is to conduct sensitivity analyses to evaluate the robustness of estimates created under an ignorable model by re-estimating these parameters under a range of assumptions about the nonignorability parameters. \citep{copas1997inference,verbeke2001sensitivity,troxel2004index,ma2005index}  

In the selection model, one describes the probability of missingness as a function of the potentially missing observation; if the missingness and the outcomes are correlated (conventionally, if a nonignorability parameter is nonzero), ignorability does not hold.  As a practical matter, it may be preferable to consider the correlation to arise from confounding --- as indeed may be the case --- in that both the outcomes and the missingness indicators are associated with a third variable.  If we can identify and measure this variable, a form of conditional independence holds that guarantees ignorability.  If we cannot, we posit a form for it and consider the consequences of nonignorability on the distributions of measurable outcomes.  Such models have long served as a basis for sensitivity analysis in observational studies, where the concern is that the treatment indicators and the potential outcomes are correlated in a way that biases standard causal analyses. \citep{cornfield1959smoking,rosenbaum1983assess,lin1998assessing,mitra2007sensitivity,ding2016sensitivity,vanderweele2017sensitivity}  In this article, we apply this approach to incomplete data.

In Section \ref{model}, we describe the model and establish the general ignorability conditions. Section \ref{response} presents a response-surface method for assessing variation of parameters of interest as a function of nonignorability parameters in a parametric model.  In Section \ref{MinNI}, we adapt Cornfield's paradigm, defining as an index of sensitivity the minimum magnitude of nonignorability that produces a designated level of bias. In Section \ref{surveydataanalysis} we apply the methods to incomplete data from a sexual behavior study.  Section \ref{extensions} covers extensions of the approach, and further discussion appears in Section \ref{discussion}.

\section{Model and methods}
\label{model}
\subsection{Model and definitions}
The data consist of an outcome variable $Y=(Y_1, \dots, Y_n)$ with corresponding vector of missingness indicators $G=(G_1,\dots, G_n)$, where $G_{i}=1$ for $Y_{i}$ observed, and $G_{i}=0$ for $Y_{i}$ missing. Assume that an unmeasured variable $U=(U_1,\ldots,U_n)$ functions as \emph{confounder} in that $Y$ and $G$ are \emph{conditionally independent} given $U$; that is, the conditional distribution of $Y$ and $G$ given $U$
has the property that, for any $u$, $f^{Y,G|U}(y,g|u)=f^{Y|U}(y|u)f^{G|U}(g|u)$ for all $y$ and $g$.
Thus, the joint density simplifies to
\begin{equation}
f^{Y,G,U}(y,g,u)= f^{U}(u)f^{Y|U}(y|u)f^{G|U}(g|u).
\label{eqn:joint}
\end{equation}

The confounding, if unmeasured or not accounted for, can induce correlation between $Y$ and $G$.  Thus although we seek to create inferences for the marginal distribution of $Y$,
\begin{equation}
f^{Y}(y)=\int f^{Y|U}(y|u)f^{U}(u)du,
\label{eqn:fygen}
\end{equation}
in fact we may be only able to observe the conditional distribution of $Y$ given $G=g$:
\begin{equation}
f^{Y|G}(y|g)=\frac{\int f^{G|U}(g|u)f^{Y|U}(y|u)f^{U}(u)du}{\int f^{G|U}(g|u)f^{Y|U}(y|u)f^{U}(u)dudy}.
\label{eqn:fycggen}
\end{equation}
To this end, we establish restrictions on the conditional distribution terms in (\ref{eqn:fycggen}) that are sufficient to guarantee ignorability, which in this context means that Equations (\ref{eqn:fygen}) and (\ref{eqn:fycggen}) are the same. Throughout, we ignore any theoretical considerations about sets of measure $0$.

\begin{theorem}
Assume that either $G \independent U$, or $Y \independent U$.  Then for any $g$ such that $f^{G}(g)>0$, the distribution ignoring the missing mechanism in (\ref{eqn:fygen}) equals the correct distribution in (\ref{eqn:fycggen}).
\label{ciiwo}
\end{theorem}

Proof.
Suppose $G\independent U$.  Then $\forall g$ with $0<f^{G}(g)<1$,
$$f^{Y|G}(y|g)=
\frac{\int f^{Y|U}(y|u)f^{G}(g)f^{U}(u)du}{\int f^{Y|U}(y|u)f^G(g)f^{U}(u)dudy}=
\frac{\int f^{Y|U}(y|u)f^{U}(u)du}{\int f^{Y|U}(y|u)f^{U}(u)dudy}=
f^{Y}(y).$$
Similarly, if $Y\independent U$, then $\forall g$ with $0<f^{G}(g)<1$,
$$f^{Y|G}(y|g)=\frac{f^{Y}(y)\int f^{G|U}(g|u)f^{U}(u)du}{\int f^{G|U}(g|u)f^{U}(u)du}=f^{Y}(y).$$

If $Y$ or $U$ is discrete, one can restate the theorem with summation substituted for integration. The ignorability condition in the theorem is stronger than MCAR because it applies to all possible missing patterns that have positive density, not just the observed missing pattern.\citep{heitjan1997ignorability,little2014statistical} In practice, the relevant conditional distribution will be the one for the outcome $y$ conditional on the observed vector of missingness indicators $\Tilde{g}$ where the sample space of $y$ will be restricted to those which could agree with $\Tilde{g}$, called \emph{$y$ consistent with $\Tilde{g}$}. Thus, we develop an alternative, weaker version of Theorem \ref{ciiwo}. 

Consider the following conditions, assuming an observed value $\Tilde{g}$ of $g$:
\begin{enumerate}
\item The missingness is \emph{observed ignorable} in that for any possible $u$, $f^{G|Y,U}(\Tilde{g}|y,u)$ takes the same value for all $y$ consistent with $\Tilde{g}$.
\item $f^{G|U}(\Tilde{g}|u)$ takes the same value for all $u$.
\item
For any $y$ consistent with $\Tilde{g}$, $f^{Y|U}(y|u)$ takes the same value for all $u$.
\end{enumerate}

This leads to the following theorem:
\begin{theorem}
Under Assumption 1 and either of Assumptions 2 or 3, $f^{Y}(y)=f^{Y|G}(y|\Tilde{g})$ for all $y$ consistent with $\Tilde{g}$.
\label{ciiweak}
\end{theorem}

In practice, there may also be completely measured predictors.  In such a case the theorems go through with appropriate conditioning, as shown in Section \ref{withcovariates}. These theorems offer the simplest general ignorability conditions for the confounding model.  Ignorability is generally not testable because $U$ is typically hypothetical and $Y_i$ is available only when $G_i=1$. Our idea therefore is to define nonignorability parameters in the context of Equation (\ref{eqn:joint}), then manipulate those parameters to determine how far they must depart from the ignorable model to create a substantial difference between $f^{Y}(y)$ and $f^{Y|G}(y|\Tilde{g})$.  

\subsection{Sensitivity analysis in the confounding model}
To simplify our exposition, we consider the situation where the data represent $n$ independently and identically distributed cases, with $U$ a scalar unmeasured confounder.  Assuming that $Y$ has finite mean and variance, we base our sensitivity analysis initially on a comparison of the marginal mean of $Y$ to its mean conditional on its being observed. 

Theorems \ref{ciiwo} and \ref{ciiweak} suggest that the sensitivity parameters can represent associations between $U$ and $Y$ and between $U$ and $G$. We will describe two approaches: In the first, we depict bias conventionally by varying the nonignorability parameters over a plausible range based on a mildly parameterized model. This extends the sensitivity analysis of Rosenbaum and Rubin from confounding in observational studies to nonignorably missing data. \citep{rosenbaum1983assess}  In the second, we consider a minimally parameterized nonparametric model and define the \emph{minimum nonignorability index (MinNI)\/} to be the degree of nonignorability necessary to cause a non-negligible bias.  This is similar to the Cornfield approach to sensitivity analysis in observational research. \citep{cornfield1959smoking} As with all sensitivity analyses, ours depends in principle on the judgments of a hypothetical expert, whose role it is to identify the minimum non-negligible values of both the bias in $Y$ and the nonignorability parameters.

\section{Response-surface sensitivity analysis}
\label{response}
\subsection{Sensitivity parameters}
\label{sec:senspar}
Assume first that the confounder $U$ is binary.  Then we partially specify the joint distribution of $(Y,G,U)$ as
$$\text{Pr}[U=0]=\pi_0,$$
$$\text{Pr}[G=1|U=u]=h(\gamma_0+\gamma_1u),$$
$$\mathbb{E}[Y|U=u]=q(\beta_0+\beta_1u),$$
where $u\in\{0,1\}$ and $h(\gamma_0+\gamma_1u)$ and $q(\beta_0+\beta_1u)$ are link functions. The parameters $\pi_0$, $\gamma_1$, and $\beta_1$ describe the degree of sensitivity; they are unidentifiable because we do not observe $U$.

A standard approach to sensitivity analysis is to observe the change in a parameter of interest, in this case the marginal mean $\mathbb{E}[Y]$, as we vary the sensitivity parameters over plausible values. Under this model, the marginal and conditional means of $Y$ in terms of these sensitivity parameters are, respectively,
\begin{equation}
\mathbb{E}[Y]=q(\beta_0+\beta_1)(1-\pi_0)+q(\beta_0)\pi_0
\label{margmean}
\end{equation}
\begin{equation}
\mathbb{E}[Y|G=1]=\frac{q(\beta_0+\beta_1)h(\gamma_0+\gamma_1)(1-\pi_0)+q(\beta_0)h(\gamma_0)\pi_0}{h(\gamma_0+\gamma_1)(1-\pi_0)+h(\gamma_0)\pi_0}
\label{condmean}
\end{equation}

\subsection{Estimation of means with specified nonignorability parameters}
\label{subsec:estpar}
Under this model, $\text{Pr}[G=1]$ and $\mathbb{E}[Y|G=1]$ are directly estimable from the data as $\hat{p}$ and $\hat{\mu}_c$, respectively.  With $\pi_0$, $\gamma_1$, and $\beta_1$ fixed, and observing a random sample of $Y$ values, some of which may be missing, one can readily estimate $\gamma_0$ and $\beta_0$. \citep{rosenbaum1983assess} The first estimable term is
\begin{equation}
\text{Pr}[G=1]=h(\gamma_0+\gamma_1)(1-\pi_0)+h(\gamma_0)\pi_0.
\label{missest}
\end{equation}
Thus we have two equations ((\ref{condmean}) and (\ref{missest})) and two unknowns ($\gamma_0$ and $\beta_0$). We calculate the marginal mean in (\ref{margmean}) as follows:
\begin{enumerate}
\item Solve equation (\ref{missest}) for $\hat{\gamma}_0$, with $\hat{p}$ and $\pi_0, \gamma_1$\label{stepgamma} fixed;
\item Solve equation (\ref{condmean}) for ${\beta}_0$, with $\pi_0, \gamma_1, \beta_1$ fixed, $\hat{\gamma}_0$ from step \ref{stepgamma}, and $\hat{\mu}_c$ estimated directly from the data;
\item Substitute $\hat{\beta}_0$ and $\pi_0, \beta_1$ into (\ref{margmean}) to estimate the marginal mean.
\end{enumerate}
Appendix \ref{app:estpar} presents details for the special case where both link functions are logistic, as would be applicable with a binary outcome.

Theorems \ref{ciiwo} and \ref{ciiweak} assert that if $\gamma_1=0$ or $\beta_1=0$, there is no difference between $\mathbb{E}[Y]$ and $\mathbb{E}[Y|G=1]$.  Thus if small values of these parameters lead to substantial variation in $\mathbb{E}[Y]$, we deem the results sensitive. If the response surface for $\mathbb{E}[Y]$ as a function of the sensitivity parameters is flat, then only large values of the sensitivity parameters imply non-negligible changes in $\mathbb{E}[Y|G=1]$, and inferences are insensitive.

If the notional unmeasured covariate $U$ is other than binary, the specification of the distribution for $U$ is  more complex and may involve more parameters. The distributions of $G$ given $U$ and $Y$ given $U$ are indexed by link functions, whose specification induces an additional source of sensitivity. Thus, semiparametric or nonparametric models might be more satisfactory for this application.

\section{Identifying the minimum non-negligible nonignorability}
\label{MinNI}
The response-surface analysis directly investigates the bias by mapping the effects of nonignorability on the distribution of $Y$.  A complementary approach is to identify minimum values for the sensitivity parameters that yield a designated level of change ---  in this case, a pre-specified maximum negligible difference between $\mathbb{E}[Y|G=1]$ and $\mathbb{E}[Y]$.  We denote these parameter values \emph{MinNI}, for \emph{Min}imum \emph{N}on\emph{I}gnorability.  We seek moreover to conduct the analysis with a minimally parameterized model.

\subsection{MinNI in the difference scale}
Assume again a binary confounder $U$. We first note that
\begin{equation} 
\mathbb{E}[Y] -\mathbb{E}[Y|G=1]=  \left(\mathbb{E}[Y|G=0]-\mathbb{E}[Y|G=1]\right)\text{Pr}[G=0].
\label{meandiff}
\end{equation}
Clearly, unless $0<\text{Pr}[G=0]<1$ there is no need for a sensitivity analysis. Expanding the bias in Equation (\ref{meandiff}) in terms of the unmeasured confounder $U$, we observe that the difference between $\mathbb{E}[Y|G=0]$ and $\mathbb{E}[Y|G=1]$ can be decomposed into the product of the difference between $\mathbb{E}[Y|U=1]$ and $\mathbb{E}[Y|U=0]$ and the difference between $\text{Pr}[U=1|G=1]$ and $\text{Pr}[U=1|G=0]$.  Details appear in Appendix \ref{proofdec}.

Define the sensitivity parameters as the two differences
$$\text{ED}_{YU}=\mathbb{E}[Y|U=1]-\mathbb{E}[Y|U=0],~~
\text{RD}_{UG}=\text{Pr}[U=1|G=1]-\text{Pr}[U=1|G=0],$$
and observe that
\begin{equation}
\left|\mathbb{E}[Y]-\mathbb{E}[Y|G=1]\right| =
\left|\text{ED}_{YU}\text{RD}_{UG}\text{Pr}[G=0]\right|.
\label{eqdifdec}
\end{equation}

\begin{figure}[!ht]
\centering
  \includegraphics[width=8.5cm]{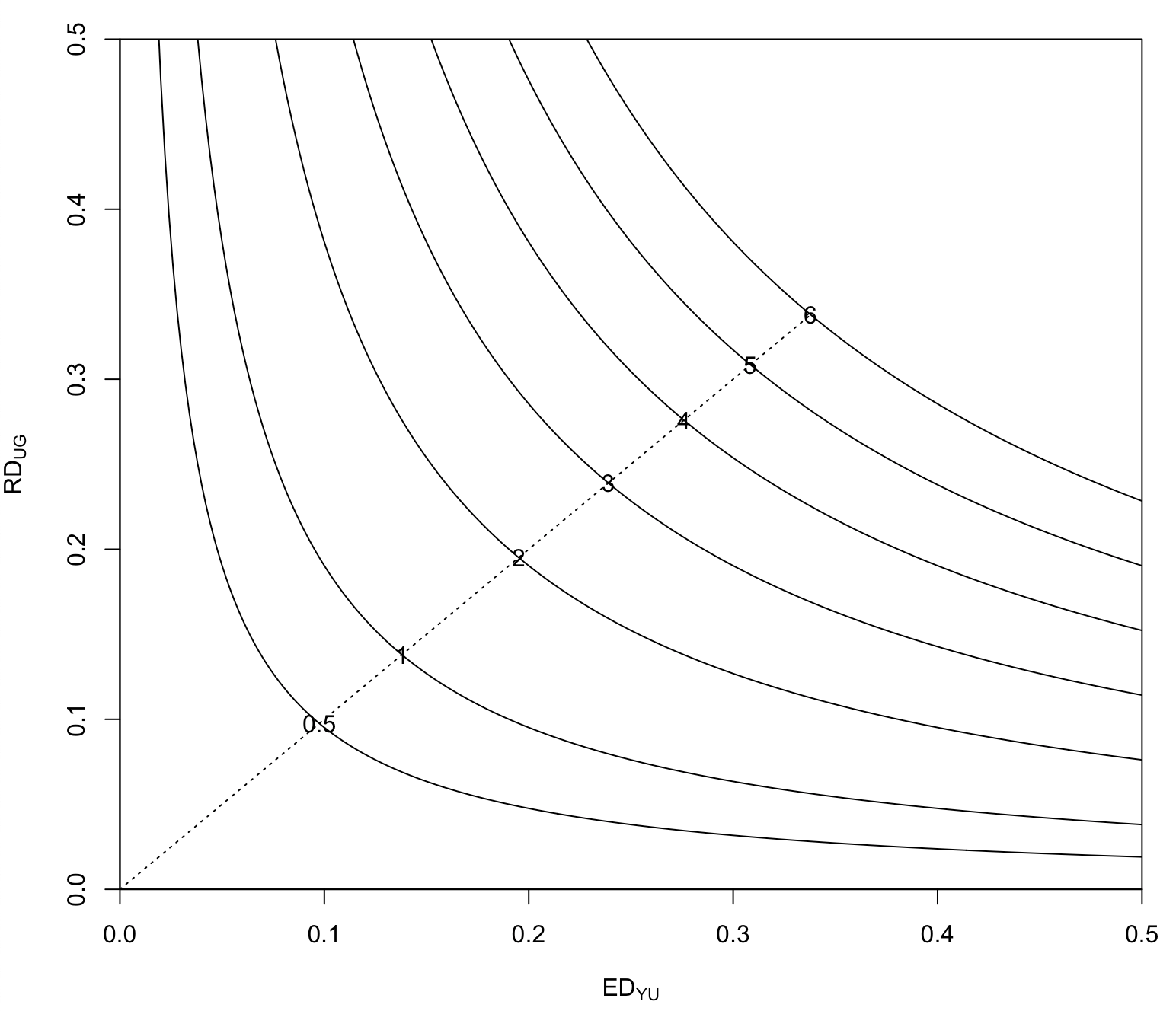}
  \caption{The equal-bias plot of $\text{ED}_{YU}$ and $\text{RD}_{UG}$ for the sexual behavior survey data. The numbers on the curves denote the bias in standard error units, with corresponding MinNI values (left to right) (0.10, 0.10), (0.14, 0.14), (0.20, 0.20), (0.24, 0.24), (0.28, 0.28), (0.31, 0.31), and (0.34, 0.34).}
  \label{fig:sbsindex}
\end{figure}

We can construct an insensitive region by specifying a maximum negligible difference for the bias as
\begin{equation}
\left|\mathbb{E}[Y]-\mathbb{E}[Y|G=1]\right| \leq k\sigma_{Y|G=1},
\label{difsc}
\end{equation}
where $\sigma_{Y|G=1}$ is the standard deviation of $Y$ given it is observed, and $k$ is a positive constant defined for the context, possibly related to sample size. From (\ref{eqdifdec}) and (\ref{difsc}), we obtain the indifference region for the nonignorable parameters to be
\begin{equation}
\left|\text{ED}_{YU}\text{RD}_{UG}\right|\leq \frac{k\sigma_{Y|G=1}}{\text{Pr}[G=0]}.
\label{difdec}
\end{equation}
The inequality (\ref{difdec}) describes the relations among the maximum tolerable change and the sensitivity parameters. To define a single index of sensitivity, we identify the combination of sensitivity parameters that satisfies this constraint and is closest to the origin.  We call this the \emph{MinNI for the mean}. For a continuous outcome, the optimization process is
\begin{flalign*}
 \text{Minimize:} &&  (\text{ED}_{YU}^2&+\text{RD}_{UG}^2)  &\\ \text{Subject to:} && \left|\text{ED}_{YU}\text{RD}_{UG}\right|&\leq \frac{k\sigma_{Y|G=1}}{\text{Pr}[G=0]};&\\ && |\text{ED}_{YU}|&\in (0,\infty);&\\ && |\text{RD}_{UG}|&\in (0,1).
\end{flalign*}
The closed-form feasible solution (i.e. MinNI) for ($|\text{ED}_{YU}|$,$|\text{RD}_{UG}|$) is
$$\left(\text{max}\left\{\frac{k\sigma_{Y|G=1}}{\text{Pr}[G=0]},\sqrt{\frac{k\sigma_{Y|G=1}}{\text{Pr}[G=0]}}\right\},\text{min}\left\{1,\sqrt{\frac{k\sigma_{Y|G=1}}{\text{Pr}[G=0]}}\right\}\right).$$
For a binary outcome, the range of $|\text{ED}_{YU}|$ in the optimization procedure is $(0,1)$, and
$$\text{MinNI} = \left(\sqrt{\frac{k\sigma_{Y|G=1}}{\text{Pr}[G=0]}},\sqrt{\frac{k\sigma_{Y|G=1}}{\text{Pr}[G=0]}}\right),$$
where $k\sigma_{Y|G=1}\leq \text{Pr}[G=0]$. If MinNI is large, the sampling inference ignoring the missing data is plausibly robust.  If it is small, ignoring the missing mechanism could cause a considerable bias.  Figure \ref{fig:sbsindex} illustrates the sensitivity analysis of the example discussed in Section \ref{surveydataanalysis} below.

\subsection{MinNI in the ratio scale}
\begin{figure}[!ht]
\centering
  \includegraphics[width=8.5cm]{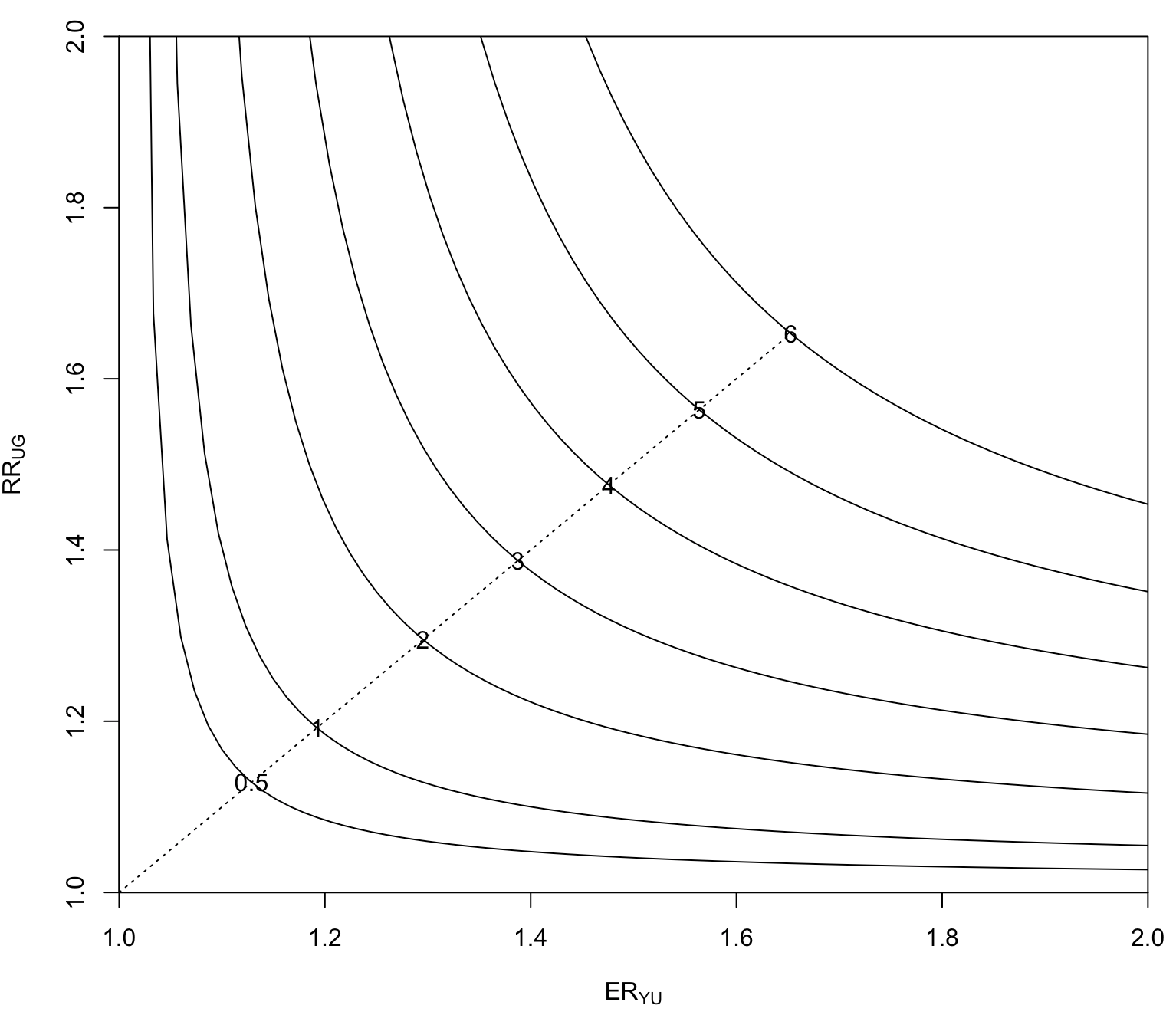}
  \caption{The equal-bias plot of $\text{ER}_{YU}$ and $\text{RR}_{UG}$ for the sexual behavior survey data. The numbers on the curves denote the bias in standard error units, with corresponding MinNI values (left to right) (1.13, 1.13), (1.19, 1.19), (1.30, 1.30), (1.39, 1.39), (1.48, 1.48), (1.56, 1.56), and (1.65, 1.65).}
  \label{fig:sbsindexratio}
\end{figure}

For categorical variables, it might be preferable to describe bias on the ratio scale. Analogously with Equation (\ref{meandiff}), we observe that
\begin{equation}
    \frac{\mathbb{E}[Y]}{\mathbb{E}[Y|G=1]}=\text{Pr}[G=1]+\text{Pr}[G=0]\frac{\mathbb{E}[Y|G=0]}{\mathbb{E}[Y|G=1]},
    \label{meanratio}
\end{equation}
where $\mathbb{E}[Y|G=1]\neq 0$. Defining the nonignorability parameters as the ratios
$$\text{ER}_{YU}=\frac{\mathbb{E}[Y|U=1]}{\mathbb{E}[Y|U=0]},~~
\text{RR}_{UG}=\frac{\text{Pr}[U=1|G=1]}{\text{Pr}[U=1|G=0]},$$
we obtain
\begin{equation}
    \frac{\mathbb{E}[Y|G=0]}{\mathbb{E}[Y|G=1]}=\frac{\text{ER}_{YU}-1+\frac{1}{\text{Pr}[U=1|G=0]}}{(\text{ER}_{YU}-1)\text{RR}_{UG}+\frac{1}{\text{Pr}[U=1|G=0]}}.
    \label{eqrsdec}
\end{equation}
Because we cannot identify $\text{Pr}[U=1|G=0]$, the best we can do is to obtain inequalities on the ratio in Equation (\ref{eqrsdec}), whose right-hand side is a monotone function of $\frac{1}{\text{Pr}[U=1|G=0]}\in (\text{RR}_{UG}, \infty)$. We express the bounding inequality for the original ratio as
\begin{equation}
\left|\frac{\mathbb{E}[Y]}{\mathbb{E}[Y|G=1]}-1\right|\leq
\left|\frac{(\text{ER}_{YU}-1)(\text{RR}_{UG}-1)}{\text{ER}_{YU}\text{RR}_{UG}}\right|\text{Pr}[G=0],
\label{bouieqrsdec}
\end{equation}
where $\text{ER}_{YU}\in (-\infty, \infty)$ and $\text{RR}_{UG}\in (0,\infty)$. When $Y$ is binary, we can specify an indifference region on the ratio scale by dividing both sides in (\ref{difsc}) by $\mathbb{E}[Y|G=1]$ to obtain 
\begin{equation}
\left|\frac{\mathbb{E}[Y]}{\mathbb{E}[Y|G=1]}-1\right|\leq |k\text{CV}_{Y|G=1}|.
\label{rasc}
\end{equation}
Here $\text{CV}_{Y|G=1}$ is the coefficient of variation of $Y$ given that it is observed. The parameters $\text{Pr}[G=0]$, $\sigma_{Y|G=1}$ and $\text{CV}_{Y|G=1}$ are all estimable from the data. To be conservative, we make the upper bound of the ratio in inequality (\ref{bouieqrsdec}) less than the specified detectable difference from (\ref{rasc}). The indifference region for the nonignorable ratio parameters is then
\begin{equation}
\left|\frac{(\text{ER}_{YU}-1)(\text{RR}_{UG}-1)}{\text{ER}_{YU}\text{RR}_{UG}}\right|\leq \frac{|k\text{CV}_{Y|G=1}|}{\text{Pr}[G=0]}
\label{rsdec}
\end{equation}
To obtain a sensitivity index, we identify the closest point to $(1,1)$.  Assuming, without loss of generality, that both $\text{ER}_{YU}$ and $\text{RR}_{UG}$ exceed $1$, the optimization process is
\begin{flalign*}
 \text{Minimize:} &&  (\text{ER}_{YU}-1)^2&+(\text{RR}_{UG}-1)^2  &\\ \text{Subject to:} && \frac{(\text{ER}_{YU}-1)(\text{RR}_{UG}-1)}{\text{ER}_{YU}\text{RR}_{UG}}&\leq \frac{|k\text{CV}_{Y|G=1}|}{\text{Pr}[G=0]};&\\ && \text{ER}_{YU}&\in (1, \infty);&\\ && \text{RR}_{UG}&\in (1, \infty).
\end{flalign*}
The closed-form solution for ($\text{ER}_{YU}$,$\text{RR}_{UG}$) is then
$$\text{MinNI} = \left(\frac{1}{1-\sqrt{\frac{|k\text{CV}_{Y|G=1}|}{\text{Pr}[G=0]}}},\frac{1}{1-\sqrt{\frac{|k\text{CV}_{Y|G=1}|}{\text{Pr}[G=0]}}}\right),$$
where $|k\text{CV}_{Y|G=1}|< \text{Pr}[G=0]\leq 1$. The interpretation is the same as for the difference scale; see Figure \ref{fig:sbsindexratio}.

\section{Sensitivity analysis for the Edinburgh sexual behavior survey}
\label{surveydataanalysis}
\subsection{The data}
Investigators surveyed 6,136 randomly selected students at the University of Edinburgh in 1993. The parameter of main interest was the fraction responding ``yes'' to the question ``Have you ever had sexual intercourse?'', which 2,308 students (37.6\%) declined to answer. \citep{raab1999information,troxel2004index,xie2018measuring} The observed proportion of positive responses, estimating $\mathbb{E}[Y|G=1]$, is 0.7320 with standard error 0.0072. There is concern that nonresponders could have different patterns of sexual behavior compared to responders, potentially inducing a bias when estimating the parameter of interest. We describe below a sensitivity analysis for this proportion. 

\subsection{A response-surface sensitivity analysis}
\begin{figure}[!ht]
\centering
\includegraphics[width=10cm]{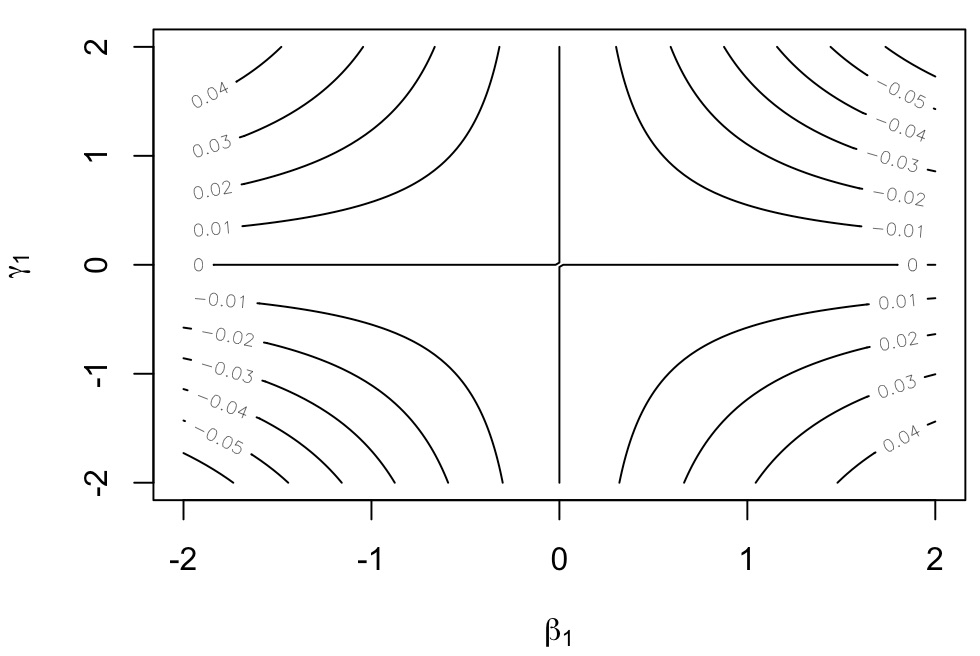}
\caption{Isobols of $\mathbb{E}[Y] - \mathbb{E}[Y|G=1]$ in terms of $\gamma_1$ and $\beta_1$, fixing $\pi_0=0.5$.}
\label{fig:sbsebp}
\end{figure}

Table \ref{tab:diftab} displays the bias as a function of the sensitivity parameters, $\pi_0$, $\beta_1$, and $\gamma_1$.  A plot of equal-bias contours in $\gamma_1$ and $\beta_1$, with $\pi_0$ fixed at $0.5$, appears in Figure \ref{fig:sbsebp}. We fixed $\pi_0=0.5$ because this value appears to give the largest bias.

\begin{table}[ht!]
\centering
\caption{$\mathbb{E}[Y] - \mathbb{E}[Y|G=1]$ as a function of the sensitivity parameters.}
\begin{tabular}{ccccc}
\hline
&&\multicolumn{3}{c}{$\pi_0$}\\
$\exp(\beta_1)$&$\exp(\gamma_1)$& 0.1&0.5&0.9\\
\hline
\multirow{2}{*}{2} & 2 & $-$0.0037 & $-$0.0088 & $-$0.0025 \\\cline{2-5}
& 3 & $-$0.0059 & $-$0.0139 & $-$0.0037 \\
\hline
\multirow{2}{*}{3} & 2 & $-$0.0061 & $-$0.0138 & $-$0.0036 \\\cline{2-5}
& 3 & $-$0.0097 & $-$0.0218 & $-$0.0053 \\
\hline
\end{tabular}
\label{tab:diftab}
\end{table}

In Table \ref{tab:diftab}, the absolute magnitude of the bias is modest as a fraction of the estimated parameter, reaching values no larger than about 3\% on a relative scale.  For purposes of statistical inference, however, the sensitivity is substantial, as the largest bias is roughly $3$ times the nominal standard error.  The equal-bias plot in Figure \ref{fig:sbsebp} indicates that moderate values of $\gamma_1$ and $\beta_1$ can lead to $2$-SE changes to the mean. The analysis thus suggests that estimation of the proportion of students who had had sexual intercourse is sensitive to nonignorability.

\subsection{A MinNI sensitivity analysis compared with ISNI analysis}
Here we set the maximum negligible bias to be $1$ standard error of the observed proportion (here $k\sigma_{Y|G=1}=0.0072$) and compute minimum values of the sensitivity parameters that produce this level of displacement. The MinNI for the difference scale, ($\text{ED}_{YU}$,$\text{RD}_{UG}$)=(0.14,0.14) from Figure \ref{fig:sbsindex} and for the ratio scale ($\text{ER}_{YU}$,$\text{RR}_{UG}$)=(1.19,1.19) from Figure \ref{fig:sbsindexratio}. The index is in both cases small, suggesting that the sampling inference for the true proportion of having sexual intercourse is sensitive. That is, even a modest disturbance from the ignorable model can induce a substantial bias into our estimate of the population proportion, rendering tests and confidence intervals for this parameter unreliable. 

We compare this analysis with an application of the likelihood-based ISNI (index of local sensitivity to nonignorability) sensitivity analysis.\citep{troxel2004index,xie2018measuring}  With ISNI, the key sensitivity statistic, denoted $c$, measures the approximate minimum standardized magnitude of nonignorability needed to induce a 1-SE change in the maximum likelihood estimate of the parameter of interest.  A value $c<1$ is generally taken as evidence of sensitivity.  For the proportion replying yes in the Edinburgh data, we compute $c=0.097$, suggesting strong sensitivity and agreeing with our frequentist analysis.

\subsection{Dependence of MinNI on the fraction of missing data}
Measures of sensitivity to nonignorability depend critically on the fraction of missing data; indeed the ISNI measure for a univariate normal mean with missing observations is proportional to the fraction missing. \citep{troxel2004index}  To illustrate this relationship, we artificially varied the fraction of missing observations while holding the observed fraction of positive responses constant. We repeated the analysis with artificial missingness fractions set to 0.1 and 0.2, both smaller than the observed value of 0.376. Table \ref{tab:artmissing} shows the dependence of the bias for $\mathbb{E}[Y]$ as a function of the response-surface sensitivity parameters.  Recalling that the standard error of the observed fraction of responses is $0.0072$, it is clear that for smaller fractions of missing data, sensitivity is modest except for the most extreme levels of confounding.
\begin{table}[ht!]
\centering
\caption{$\mathbb{E}[Y] - \mathbb{E}[Y|G=1]$ as a function of the sensitivity parameters with $\pi_0=0.5$.}
\begin{tabular}{ccccc}
\hline
&&\multicolumn{3}{c}{Fraction missing}\\
$\exp(\beta_1)$&$\exp(\gamma_1)$& 0.1&0.2&0.376\\
\hline
\multirow{2}{*}{2} & 2 & $-$0.0023 & $-$0.0046 &  $-$0.0088
\\\cline{2-5}
& 3 & $-$0.0035 & $-$0.0071 & $-$0.0139 \\
\hline
\multirow{2}{*}{3} & 2 & $-$0.0035 & $-$0.0072 & $-$0.0138 
\\\cline{2-5}
& 3 & $-$0.0054 & $-$0.0111 & $-$0.0218 \\
\hline
\end{tabular}
\label{tab:artmissing}
\end{table}

Table \ref{tab:artMinNI} shows MinNI values for the difference and ratio scale sensitivity analyses under the alternative fractions of missing observations.  The interpretation of these values is that one would require weaker levels of confounding to induce a non-negligible bias in the observed fraction of positive responses.
\begin{table}[ht!]
\centering
\caption{The MinNI giving one standard error bias with different fractions of missing data.}
\begin{tabular}{cccc}
\hline
&\multicolumn{3}{c}{Fraction missing}\\
Scale& 0.1&0.2&0.376\\
\hline
($|\text{ED}_{YU}|$,$|\text{RD}_{UG}|$) & (0.27,0.27) & (0.19,0.19) & (0.14,0.14) \\
\hline
($\text{ER}_{YU}$,$\text{RR}_{UG}$)& (1.46,1.46) & (1.28,1.28) & (1.19,1.19) \\
\hline
\end{tabular}
\label{tab:artMinNI}
\end{table}

\section{Some extensions of the basic sensitivity analysis}
\label{extensions}
\subsection{A categorical confounder}
It is straightforward to extend our analysis to the case of an unmeasured confounder with $m>2$ levels.  For the difference scale, denote the confounding relations as follows: 
$$\text{MD}_{YU}=\max_i\mathbb{E}[Y|U=u_i]-\min_i\mathbb{E}[Y|U=u_i],$$
$$\text{MD}_{UG}=\max_i\left[\text{Pr}\left[U=u_i|G=1\right]-\text{Pr}\left[U=u_i|G=0\right]\right].$$
In Appendix \ref{app:boucate} we derive the bounding inequality to be
\begin{align} 
\left|\mathbb{E}[Y]-\mathbb{E}[Y|G=1]\right| &\leq \left|(m-1)\text{MD}_{YU}\text{MD}_{UG}\text{Pr}[G=0]\right|.
\label{muldifdec}
\end{align}
To be conservative, we make the upper bound of the difference less than $k$ standard deviations of the observed standard deviation $\sigma_{Y|G=1}$,
\begin{equation}
\left|\text{MD}_{YU}\text{MD}_{UG}\right|\leq \frac{k\sigma_{Y|G=1}}{(m-1)\text{Pr}[G=0]}.
\label{muldifind}
\end{equation}
The dependence of the sensitivity on the number of categories is the same as found in Ding and VanderWeele (2014). 

For the relative ratio scale, we denote the confounding parameters to be
$$\text{ER}_{YU(i)}=\frac{\mathbb{E}[Y|U=u_i]}{\underset{i}{\text{min}} \mathbb{E}[Y|U=u_i]},~~
\text{RR}_{UG(i)}=\frac{\text{Pr}[U=u_i|G=1]}{\text{Pr}[U=u_i|G=0]},$$
$$\text{MR}_{YU}=\max_i \text{ER}_{YU(i)},~~\text{MR}_{UG}=\max_i \text{RR}_{UG(i)}.$$
Without loss of generality, we can take all of these parameters to be greater than 1. In Appendix \ref{app:boucate} we show the bounding inequality to be
\begin{equation} 
\left|\frac{\mathbb{E}[Y]}{\mathbb{E}[Y|G=1]}-1\right| 
\leq \left|\frac{(\text{MR}_{YU}-1)(\text{MR}_{UG}-1)}{\text{MR}_{YU}\text{MR}_{UG}}\right|.
\label{mulradec}
\end{equation}
This leads to the conservative indifference region
\begin{equation}
\frac{(\text{MR}_{YU}-1)(\text{MR}_{UG}-1)}{\text{MR}_{YU}\text{MR}_{UG}} \leq \frac{k\text{CV}_{Y|G=1}}{\text{Pr}[G=0]}.
\label{mulraind}
\end{equation}
The corresponding MinNI derivations appear in Appendix \ref{app:senscate}.

\subsection{Sensitivity analysis for the variance}
\label{subsec:variance}
So far we have only considered bias in the mean of $Y$, but bias can also affect the variance. In an obvious notation, we define $\sigma_B^2$ to be the variance of a random variable $B$, potentially with conditioning.  Setting
$$\text{VD}_{YU}=\sigma^2_{Y|U=0}-\sigma^2_{Y|U=1}, \text{VD}_{UG}=\sigma^2_{U|G=0}-\sigma^2_{U|G=1},$$
we obtain
\begin{equation}
  \sigma^2_Y-\sigma^2_{Y|G=1}=\left\{\text{VD}_{YU}\text{RD}_{UG}+\text{ED}_{YU}^2\text{VD}_{UG}+\text{ED}_{YU}^2\text{RD}_{UG}^2\text{Pr}[G=1]\right\}\text{Pr}[G=0].
  \label{variancedecom}
\end{equation}
Theorem \ref{ciiwo} asserts that if $G \independent U$ or $Y \independent U$, then the difference in Equation (\ref{variancedecom}) is 0. For the comparison of means, if either $\text{ED}_{YU}$ or $\text{RD}_{UG}$ is 0, there is no bias, but for the comparison of variance, this condition is not sufficient because $\text{VD}_{YU}$ or $\text{VD}_{UG}$ might not be 0. Commonly, the main moment of interest is the mean and it is shown that the first-order Taylor expansion of $\sigma^2_{Y|G=1}$ is equal to $\sigma^2_Y$. \citep{troxel2004index,ma2004sensitivity} We can readily derive analogous results for estimating the conditional distribution of $Y$ given $X$.

\subsection{Analysis with completely measured covariates}
\label{withcovariates}
Many studies will include a list of baseline variables that, if unobserved, would confound the association of outcome and missingness; we denote such variables $X$. We can readily generalize Theorems \ref{ciiwo} and \ref{ciiweak} to cover estimation of the distribution of $Y$ given $X$. 
\begin{theorem}
Assume that $Y\independent G|(X,U)$ and that either $G \independent U|X$ or $Y \independent U|X$.  Then for any $g$ and $x$ such that $f^{G,X}(g,x)>0$, the distribution ignoring the missing mechanism $f^{Y|X}(y|x)$, equals the correct distribution $f^{Y|G,X}(y|g,x)$.
\label{ciicond}
\end{theorem}
To generalize Theorem \ref{ciiweak}, we define the following assumptions, assuming that $\Tilde{g}$ is the observed value of $G$:
\begin{enumerate}
\item The missingness is \emph{observed ignorable} in that for any possible $u$ and $x$, $f^{G|Y,X,U}(\Tilde{g}|y,x,u)$ takes the same value for all $y$ consistent with $\Tilde{g}$.
\item For any possible $x$, $f^{G|X,U}(\Tilde{g}|x,u)$ takes the same value for all $u$.
\item
For any possible $x$ and any $y$ consistent with $\Tilde{g}$, $f^{Y|X,U}(y|x,u)$ takes the same value for all $u$.
\end{enumerate}
\begin{theorem}
Under Assumption 1 and either of Assumptions 2 or 3, $f^{Y|X}(y|x)=f^{Y|X,G}(y|x,\Tilde{g})$ for all $y$ consistent with $\Tilde{g}$.
\label{ciiweakcond}
\end{theorem}
Our analysis also readily generalizes to this situation; that is, by further conditioning on $X$ one can elucidate sensitivity as we have done above. Assume that the measured covariates $X$ are discrete. For the difference scale, denote the two confounding relations as
$$\text{ED}_{YU(X)}=\mathbb{E}[Y|X,U=1]-\mathbb{E}[Y|X,U=0],$$
$$\text{RD}_{UG(X)}=\text{Pr}[U=1|X,G=1]-\text{Pr}[U=1|X,G=0].$$
Therefore,
\begin{align*}
    |\mathbb{E}[Y|X]-\mathbb{E}[Y|X,G=1]|=|\text{ED}_{YU(X)}\text{RD}_{UG(X)}\text{Pr}[G=0|X]|.
\end{align*}
The above formula is similar to Equation (\ref{eqdifdec}). However, for the ratio scale, one naive analysis will be shown below.
$$\frac{\mathbb{E}[Y|X]}{\mathbb{E}[Y|X,G=1]}=\text{Pr}[G=1|X]+\frac{\mathbb{E}[Y|X,G=0]}{\mathbb{E}[Y|X,G=1]}\text{Pr}[G=0|X],$$
and we denote the relative ratios as
$$\text{ER}_{YU(X)}=\frac{\mathbb{E}[Y|X,U=1]}{\mathbb{E}[Y|X,U=0]},$$
$$\text{RR}_{UG(X)}=\frac{\text{Pr}[U=1|X,G=1]}{\text{Pr}[U=1|X,G=0]}.$$
Hence, 
\begin{align*}
\frac{\mathbb{E}[Y|X,G=0]}{\mathbb{E}[Y|X,G=1]}=\frac{\text{ER}_{YU(X)}-1+\frac{1}{\text{Pr}[U=1|X,G=0]}}{(\text{ER}_{YU(X)}-1)\text{RR}_{UG(X)}+\frac{1}{\text{Pr}[U=1|X,G=0]}},
\end{align*}
which recalls Equation (\ref{eqrsdec}). All the other derivations follow directly. The total discrepancy between the marginal mean and the conditional mean after adjusting for the $X$ could be the summation of the discrepancy weighted by $X$. 

\section{Discussion}
\label{discussion}
We have used a model that frames nonignorability as a consequence of unobserved confounding to devise a simple, general paradigm for sensitivity analysis in frequentist inference of incomplete data. The interpretation of nonignorability in our model mirrors the methods for analysis of sensitivity to nonignorable confounding of Cornfield, Rosenbaum and Rubin, and Ding and VanderWeele.\citep{cornfield1959smoking,rosenbaum1983assess,ding2014generalized}

Our sensitivity analysis involves identifying the minimum degree of nonignorability that causes a designated discrepancy in some comparison of $f^Y(y)$ and $f^{Y|G}(y|g)$.  In the analyses we demonstrate here, we assume that the outcomes $Y$ represent an i.i.d.~sample from some distribution, but that is not essential to the method. Moreover we have derived equations for comparisons of means and variances, but it should be possible to extend the analysis to other functionals. To be conservative, we make use of the partial identification region on the ratio scale to propose the MinNI under the most sensitive case so as to summarize the degree of nonignorability. 

When there is a continuous measured covariate $X$, it would be natural to apply parametric sensitivity models that require specification of a distributional form for $X$. \citep{lin1998assessing,mitra2007sensitivity} If the dimension of $X$ is large, the nonparametric analysis of MinNI described in Section \ref{withcovariates} will be difficult because of the sparsity of data (i.e. the curse of dimensionality). Thus, other nonignorable models based on selection specification, such as ISNI analysis or the propensity score matching approach might be much simpler. \citep{troxel2004index,ma2005index,zhang2006simple,zhao2019sensitivityvalue,zhao2019sensitivity} Under parametric sensitivity analysis, one possible extension should consider sensitivity to parametric model misspecification, for instance in link functions relating $Y$ and $G$ to $U$.\citep{gustafson2001measuring}

Our model differs from many prior developments in this area in referring to frequentist rather than likelihood/Bayesian estimation.\citep{heitjan1991ignorability,heitjan1994ignorability,heitjan1997ignorability}  In principle, we could construct a similar analysis replacing the selection specification with the confounding specification as the incompleteness mechanism in a model-based analysis, and evaluating, say, the minimum nonignorability needed to deflect maximum likelihood estimates by a designated amount.  This would render our approach comparable to the ISNI sensitivity analysis.

\nocite{*}
\bibliographystyle{apa}
\bibliography{sauc}

\section{Appendix}\label{appendix}
\subsection{Estimation of the unknown parameters with a binary outcome}
\label{app:estpar}
Assume a population of $N$ units, where for each unit $i$ there is an associated vector $(Y_i, G_i)\in \{0,1\}\times \{0,1)\}$ for the outcome of interest and its corresponding missingness indicator. The parameter of interest is
$\text{Pr}[Y=1]$.  In Section \ref{sec:senspar} we assume the two link functions to be logit functions.  Suppose observing a random sample of $Y$ values of size $n$ with $n_m$ missing and $n_o$ observed, where $n=n_m+n_o$. Then, the maximum likelihood estimates(MLEs) for the probability of missing and $\text{Pr}[Y=0|G=1]$ can be estimated by the proportion of missing in the sample, $$\hat{p}_m=\frac{n_m}{n}$$ and $$\hat{\mu}_c=1-\frac{1}{n_o}\sum_{i:g_i=1}^ny_i.$$ Following the first two steps in section \ref{subsec:estpar} with fixed $\pi_0, \gamma_1, \beta_1$, 
\begin{equation}
    \text{Pr}[G=0]=\frac{\pi_0}{1+\exp(\gamma_0)}+\frac{1-\pi_0}{1+\exp(\gamma_0+\gamma_1)}.
    \label{estgm0}
\end{equation}
Plug $\hat{p}_m$ into equation (\ref{estgm0}) with fixed $\pi_0, \gamma_1$ to obtain
$$\exp(\hat{\gamma}_0)=\frac{-[(\hat{p}_m-\pi_0)\exp(\gamma_1)+\hat{p}_m+\pi_0-1]+\sqrt{[(\hat{p}_m-\pi_0)\exp(\gamma_1)+\hat{p}_m+\pi_0-1]^2+4\exp(\gamma_1)\hat{p}_m(1-\hat{p}_m)}}{2\exp(\gamma_1)\hat{p}_m}.$$
Then
\begin{equation}
    \text{Pr}[Y=0|G=1]=\frac{w}{1+\exp(\beta_0)}+\frac{1-w}{1+\exp(\beta_0+\beta_1)},
    \label{estbet0}
\end{equation}
where
\begin{align*}
    w=\text{Pr}[U=0|G=1]&=\frac{\text{Pr}[G=1|U=0]\text{Pr}[U=0]}{\text{Pr}[G=1|U=0]\text{Pr}[U=0]+\text{Pr}[G=1|U=1]\text{Pr}[U=1]}\\
    &=\pi_0\left[\pi_0+\frac{\exp(\gamma_1)[1+\exp(\gamma_0)]}{1+\exp(\gamma_0+\gamma_1)}(1-\pi_0)\right]^{-1},
\end{align*}
and substituting $\pi_0, \gamma_1$ and $\hat{\gamma}_0$ to estimate $w$, denoted as $\hat{w}$. Plugging $\hat{w}$ and $\hat{\mu}_c$ into Equation (\ref{estbet0}) gives
$$\exp(\hat{\beta}_0)=\frac{-[(\hat{\mu}_c-\hat{w})\exp(\beta_1)+\hat{\mu}_c+\hat{w}-1]+\sqrt{[(\hat{\mu}_c-\hat{w})\exp(\beta_1)+\hat{\mu}_c+\hat{w}-1]^2+4\exp(\beta_1)\hat{\mu}_c(1-\hat{\mu}_c)}}{2\exp(\beta_1)\hat{\mu}_c}.$$

\subsection{Proof of Equation \ref{eqdifdec}}
\label{proofdec}
\begin{align*}
    \mathbb{E}[Y|G=0]-\mathbb{E}[Y|G=1] &=\mathbb{E}[Y|U=1,G=0]\text{Pr}[U=1|G=0]+\mathbb{E}[Y|U=0,G=0]\text{Pr}[U=0|G=0]\\
    &-\left\{\mathbb{E}[Y|U=1,G=1]\text{Pr}[U=1|G=1]+\mathbb{E}[Y|U=0,G=1]\text{Pr}[U=0|G=1]\right\}\\
    &=\mathbb{E}[Y|U=1]\text{Pr}[U=1|G=0]+\mathbb{E}[Y|U=0](1-\text{Pr}[U=1|G=0])\\
    &-\mathbb{E}[Y|U=1]\text{Pr}[U=1|G=1]-\mathbb{E}[Y|U=0](1-\text{Pr}[U=0|G=1])\\
    &=\mathbb{E}[Y|U=1](\text{Pr}[U=1|G=0]-\text{Pr}[U=1|G=1])\\
    &-\mathbb{E}[Y|U=0](\text{Pr}[U=1|G=0]-\text{Pr}[U=1|G=1])\\
    &=(\mathbb{E}[Y|U=1]-\mathbb{E}[Y|U=0])(\text{Pr}[U=1|G=0]-\text{Pr}[U=1|G=1])\\
\end{align*}

\subsection{A categorical confounder}
\subsubsection{Bounding inequality derivation with categorical confounder}
\label{app:boucate}
For the difference scale, $\text{MD}_{YU}$ and $\text{MD}_{UG}$ have been defined in the same notation from the above section. Without loss of generality, define level $m$ of $U$ to minimize $\mathbb{E}[Y|U=u_j]$. To simplify the derivation, define $\left|\mathbb{E}[Y|G=0]-\mathbb{E}[Y|G=1]\right|$ as $D$. It has been derived that $\left|\mathbb{E}[Y]-\mathbb{E}[Y|G=1]\right|=D\times\text{Pr}[G=0]$ and the only formula required to be derived in terms of the two defined relations is $D$.
\begin{align*}
D&=\left|\sum_{i=1}^{m}\mathbb{E}[Y|U=u_i]\text{Pr}[U=u_i|G=0]-\sum_{i=1}^{m}\mathbb{E}[Y|U=u_i]\text{Pr}[U=u_i|G=1]\right|\\
&=\left|\sum_{i=1}^{m-1}\mathbb{E}[Y|U=u_i]\text{Pr}[U=u_i|G=0]+\mathbb{E}[Y|U=u_m][1-\sum_{i=1}^{m-1}\text{Pr}[U=u_i|G=0]]\right.\\
&\left.-\{\sum_{i=1}^{m-1}\mathbb{E}[Y|U=u_i]\text{Pr}[U=u_i|G=1]+\mathbb{E}[Y|U=u_m][1-\sum_{i=1}^{m-1}\text{Pr}[U=u_i|G=1]]\}\right|\\
&=\left|\sum_{i=1}^{m-1}[\mathbb{E}[Y|U=u_i]-\mathbb{E}[Y|U=u_m]]\text{Pr}[U=u_i|G=0]\right.\\
&\left.-\sum_{i=1}^{m-1}[\mathbb{E}[Y|U=u_i]-\mathbb{E}[Y|U=u_m]]\text{Pr}[U=u_i|G=1]\right|\\
&\leq\left|\text{MD}_{YU}\sum_{i=1}^{m-1}[\text{Pr}[G=u_i|G=0]-\text{Pr}[G=u_i|G=1]]\right|\\
&\leq |(m-1)\text{MD}_{YU}\text{MD}_{UG}|
\end{align*}
This is inequality (\ref{muldifdec}).

For the ratio scale, $\text{ER}_{YU(i)}$, $\text{RR}_{UG(i)}$, $\text{MR}_{YU}$ and $\text{MR}_{UG}$ have been defined in the same way with all of them assumed to be greater than 1. Without loss of generality, let
$$\mathbb{E}[Y|U=u_1]=\text{min}_i\mathbb{E}[Y|U=u_i],~~
\mathbb{E}[Y|U=u_m]=\text{max}_i\mathbb{E}[Y|U=u_i].$$
Denote $R$ as $\frac{\mathbb{E}[Y|G=0]}{\mathbb{E}[Y|G=1]}$. Then
\begin{align*}
R&=\frac{\sum_{i=1}^m\mathbb{E}[Y|U=u_i]\text{Pr}[G=u_i|G=0]}{\sum_{i=1}^m\mathbb{E}[Y|U=u_i]\text{Pr}[G=u_i|G=1]}\\
&=\frac{\text{MR}_{YU}\text{Pr}[U=u_1|G=0]+\sum_{i=2}^{m-1}\text{ER}_{YU(i)}\text{Pr}[G=u_i|G=0]+\text{Pr}[U=u_m|G=0]}{\text{MR}_{YU}\text{Pr}[U=u_1|G=1]+\sum_{i=2}^{m-1}\text{ER}_{YU(i)}\text{Pr}[G=u_i|G=1]+\text{Pr}[U=u_m|G=1]}\\
&=\frac{\sum_{i=1}^{m-1}(\text{ER}_{YU(i)}-1)\text{Pr}[G=u_i|G=0]+1}{\sum_{i=1}^{m-1}(\text{ER}_{YU(i)}-1)\text{Pr}[G=u_i|G=1]+1}\\
&=\frac{(\text{MR}_{YU}-1)\text{Pr}[U=u_1|G=0]+\sum_{i=2}^{m-1}(\text{RR}_{YU(i)}-1)\text{Pr}[G=u_i|G=0]+1}{(\text{MR}_{YU}-1)\text{RR}_{UG(1)}\text{Pr}[U=u_1|G=0]+\sum_{i=2}^{m-1}(\text{RR}_{YU(i)}-1)\text{RR}_{UG(i)}\text{Pr}[G=u_i|G=0]+1}\\
&\geq \frac{(\text{MR}_{YU}-1)\text{Pr}[U=u_1|G=0]+1}{(\text{MR}_{YU}-1)\text{RR}_{UG(1)}\text{Pr}[U=u_1|G=0]+1}\\
&\geq \frac{\text{MR}_{YU}+\text{RR}_{UG(1)}-1}{\text{MR}_{YU}\text{RR}_{UG(1)}}\\
&\geq \frac{\text{MR}_{YU}+\text{MR}_{UG}-1}{\text{MR}_{YU}\text{MR}_{UG}}.
\end{align*}
With all the relative ratios greater than 1, $R<1$:
$$\frac{\text{MR}_{YU}+\text{MR}_{UG}-1}{\text{MR}_{YU}\text{MR}_{UG}}\leq R \leq 1$$
We can restate the bounds as
$$
\left|R-1\right|\leq \frac{(\text{MR}_{YU}-1)(\text{MR}_{UG}-1)}{\text{MR}_{YU}\text{MR}_{UG}},
$$
giving inequality (\ref{mulradec}). When some of the relative ratios are greater than 1 and others are less than 1, the derivation is more complicated, but similar tricks of finding a bounding value for the ratio of $\mathbb{E}[Y]$ and $\mathbb{E}[Y|G=1]$ could be considered.

\subsubsection{MinNI for a categorical confounder}
\label{app:senscate}
The MinNI on the risk difference scale from inequality (\ref{muldifind}) for ($\text{MD}_{YU}$,$\text{MD}_{UG}$) is, for a continuous outcome,
$$\left(\max\left\{\frac{k\sigma_{Y|G=1}}{(m-1)\text{Pr}[G=0]},\sqrt{\frac{k\sigma_{Y|G=1}}{(m-1)\text{Pr}[G=0]}}\right\}, \min\left\{1,\sqrt{\frac{k\sigma_{Y|G=1}}{(m-1)\text{Pr}[G=0]}}\right\}\right),$$ but for a binary outcome,
$$\left(\sqrt{\frac{k\sigma_{Y|G=1}}{(m-1)\text{Pr}[G=0]}},\sqrt{\frac{k\sigma_{Y|G=1}}{(m-1)\text{Pr}[G=0]}}\right),$$
where $k\sigma_{Y|G=1}\leq (m-1)\text{Pr}[G=0]$.
The MinNI on the risk ratio scale from inequality (\ref{mulraind}) for ($\text{MR}_{YU}$,$\text{MR}_{UG}$) is
$$\left(\frac{1}{1-\sqrt{\frac{k\text{CV}_{Y|G=1}}{\text{Pr}[G=0]}}},\frac{1}{1-\sqrt{\frac{k\text{CV}_{Y|G=1}}{\text{Pr}[G=0]}}}\right),$$
where $0 \leq k\text{CV}_{Y|G=1}< \text{Pr}[G=0]\leq 1$, which is the same as the binary case.
\end{document}